\documentclass[aps,prb,twocolumn,superscriptaddress]{revtex4}
\usepackage{color}
\usepackage{graphicx}
\usepackage{amsmath}
\usepackage{microtype}
\usepackage{siunitx}
\usepackage{subfigure}
\usepackage{bm}
\usepackage[colorlinks=true, letterpaper=true, pdfstartview=FitV, linkcolor=red, citecolor=blue, urlcolor=blue]{hyperref}
\begin{document}
\title{The Bulk-boundary Correspondence in Non-Hermitian Hopf-link Exceptional Line Semimetals}

\author{Zhicheng Zhang}
\affiliation{Beijing National Laboratory for Condensed Matter Physics and Institute of Physics, Chinese Academy of Sciences, Beijing 100190, China}
\affiliation{School of Physical Sciences, University of Chinese Academy of Science, Beijing 100190, China}

\author{Zhesen Yang}
\affiliation{Beijing National Laboratory for Condensed Matter Physics and Institute of Physics, Chinese Academy of Sciences, Beijing 100190, China}
\affiliation{School of Physical Sciences, University of Chinese Academy of Science, Beijing 100190, China}

\author{Jiangping Hu}\email{jphu@iphy.ac.cn}
\affiliation{Beijing National Laboratory for Condensed Matter Physics and Institute of Physics, Chinese Academy of Sciences, Beijing 100190, China}
\affiliation{CAS Center of Excellence in Topological Quantum Computation and Kavli Institute of Theoretical Sciences,
	University of Chinese Academy of Sciences, Beijing 100190, China}
\affiliation{Collaborative Innovation Center of Quantum Matter, Beijing 100871, China}

\begin{abstract}
We consider a 3-dimensional (3D) non-Hermitian exceptional line semimetal model and take open boundary conditions in x, y, and z directions separately.  In each case, we calculate the parameter regions where the bulk-boundary correspondence is broken. The breakdown of the bulk-boundary correspondence is manifested by the deviation from unit circles of generalized Brillouin zones (GBZ) and the discrepancy between spectra calculated with open boundary conditions (OBC) and periodic boundary conditions (PBC). The consistency between OBC and PBC spectra can be recovered if the PBC spectra are calculated with GBZs. We use  both unit-circle Brillouin zones (BZ) and GBZs to plot the topological phase diagrams.  The systematic analysis about the differences between the two phase diagrams suggests that it is necessary to use GBZ to characterize the bulk-boundary correspondence of non-Hermitian models.%We also propose a sufficient condition to determine whether the two phase diagrams are exactly the same. 
 %We also prove that if the solutions of the characteristic equation with $E=0$ satisfy $|\beta|=1$, the two phase diagrams will be the same.
\end{abstract} 


\maketitle

\section{\label{s:intro}Introduction}
The Hamiltonians in standard quantum mechanics are required to be Hermitian \cite{sakurai2014modern}. Over the past two decades, topological properties of Hermitian systems have been studied intensively \cite{RevModPhys.82.3045,RevModPhys.83.1057,Moore:2010aa}, such as topological insulators \cite{RevModPhys.82.3045,PhysRevLett.98.106803,PhysRevB.76.045302,PhysRevB.82.045122,Zhang:2009aa}, topological superconductors \cite{RevModPhys.83.1057,PhysRevB.78.195125,PhysRevB.82.184516,Sato_2017} and topological semimetals \cite{PhysRevB.83.205101,Burkov:2016aa,Fang_2016,PhysRevB.99.075130}. Non-zero topological invariants  always require the existence of corresponding boundary states. This fact is known to be the distinguished bulk-boundary correspondence in Hermitian systems \cite{bernevig2013topological}. 

Recently, the study of  topological states has been extended to non-Hermitian systems\cite{PhysRevLett.120.146402,yao2018edge,PhysRevLett.123.170401,PhysRevLett.123.246801,PhysRevLett.121.136802,PhysRevX.8.031079,PhysRevLett.124.086801,PhysRevLett.121.026808,YOAHIDA2019,PhysRevB.99.245116,PhysRevB.99.235112,PhysRevX.9.041015,PhysRevB.84.205128,PhysRevLett.116.133903,PhysRevLett.118.040401,PhysRevA.97.052115,PhysRevB.97.075128,PhysRevLett.118.045701,Ke:17,PhysRevA.92.012116}. Except those with PT symmetry \cite{PhysRevLett.80.5243}, the eigenvalues of non-Hermitian Hamiltonians, including    open systems \cite{Rotter_2009,malzard2015topologically,choi2010quasieigenstate,lee2014heralded}, systems with gain and loss \cite{makris2008beam,longhi2009bloch,klaiman2008visualization,Regensburger:2012aa,PhysRevLett.108.024101,Ruter:2010aa,PhysRevLett.106.213901,Feng:2012aa,PhysRevLett.103.093902,PhysRevLett.108.173901,Peng328,Fleury:2015aa,Chang:2014aa,Hodaei:2017aa,Hodaei975,Feng972,Gao:2015aa,Xu:2016aa,Ashida:2017aa,PhysRevLett.119.190401,Chen:2017aa,PhysRevX.6.021007,PhysRevB.95.125426}, and interacting electron systems where the self-energy introduced by interactions is treated as non-Hermitian terms \cite{kozii2017nonhermitian,PhysRevB.98.035141}, are generally not real. The complex eigenvalues result in novel properties in non-Hermitian systems like exceptional points and enriched topological classifications \cite{PhysRevX.8.031079,PhysRevLett.123.066405,bessho2019topological,PhysRevB.99.125103,PhysRevB.100.144106,PhysRevB.99.121101,PhysRevLett.124.056802,PhysRevLett.124.186402,yang2019fermion}.

One interesting property in non-Hermitian systems is the skin effect\cite{Xiong_2018,yoshida2019mirror,PhysRevB.99.201103,hofmann,yi2020nonhermitian,PhysRevLett.124.086801} which  states that all the eigenstates with open boundary conditions (OBC) can be localized at one side of the lattice.  Unlike the extended Bloch states in Hermitian cases \cite{yao2018edge},    the emergence of these skin modes in non-Hermitian cases indicates the breakdown of the bulk-boundary correspondence. A theorem has been proposed to determine whether there are skin modes and  whether the bulk-boundary correspondence of a non-Hermitian system is broken \cite{zhang2019correspondence}.  It claims that the nonzero winding number of periodic boundary conditions (PBC) spectrum $\nu_{E}$ with respect to any reference energy $E_{b}$ on the complex energy plane requires the existence of skin modes in corresponding OBC system, and vice versa \cite{zhang2019correspondence}.  It has also been suggested that the bulk-boundary correspondence in non-Hermitian systems may be captured by using generalized Brillouin zones (GBZ)s instead of the normal Brillouin zones \cite{yao2018edge}. 

In the GBZ approach, recovering the bulk-boundary correspondence is achieved by extending Bloch wave vector $k$ to the complex plane \cite{yao2018edge}. A systematic procedure has been proposed  to calculate GBZ numerically \cite{yokomizo2019non,zhang2019correspondence}. Specifically,  the GBZ can be obtained by solving the characteristic equation $\det[H(\beta)-E]=0$. If $\det[H(\beta)-E]$ is an irreducible algebraic polynomial of $E$ and $\beta$, the condition $|\beta_{p}|=|\beta_{p+1}|$ can lead to the GBZ, where $p$ is the order of the pole of the characteristic equation. On the basis of GBZ, the winding number $w$ is redefined in appendix A. With the redefined winding number, the parameter region corresponding to  edge modes can be predicted \cite{yao2018edge,yokomizo2019non}. More details about this method is attached in Appendix A.  Except the numerical method mentioned above, there is an analytic method that can give the explicit expression of GBZs. Algebraic GBZ equation can be translated to a geometric condition by defining auxiliary generalized Brillouin zones (aGBZ) \cite{yang2019auxiliary}. The aGBZs are calculated analytically with the help of the mathematical concept ``resultant".  We can obtain the real GBZ from aGBZs with the condition $|\beta_{p}|=|\beta_{p+1}|$. The details of this analytic method are given in Appendix B.  

A non-Hermitian generalization of nodal line semimetal is called exceptional line semimetal, which can exhibit  properties absent in the Hermitian case\cite{PhysRevA.98.042114,PhysRevB.100.054109,lee2018tidal,Li2020emergence}.   For example,  topological properties in Hermitian nodal line semimetals are protected by symmetries \cite{Fang_2016}, while non-Hermitian nodal line semimetals are not. The latter also exhibits  Hopf-link exceptional lines  in certain parameter regions \cite{yang2019non}.   The bulk-boundary correspondence and the corresponding GBZ approach has  been well studied for 1D non-Hermitian models.  However, there is not much study about 3D models. Based on these motivations, in this paper we will take OBC in x, y, and z directions separately to study the bulk-boundary correspondence of 3D non-Hermitian Hopf-link exceptional line semimetals.

\par  The paper is organized as follows. In section II, we introduce the exceptional line semimetal models.  In section III,  we study the non-Hermitian semimetal model with OBC in z direction. We derive the analytic expression of the GBZs and give the parameter regions where the bulk-boundary correspondence is broken. We manifest the breakdown of the bulk-boundary correspondence by showing the deviation from unit circles of the GBZs and  the discrepancy between spectra calculated with OBC and PBC. We also point out that the consistency between OBC and PBC spectra can be recovered if the PBC spectra are calculated with GBZs.  Finally,  we plot the topological phase diagrams calculated with both unit-circle Brillouin zones (BZs) and GBZs and discuss the reasons for the differences between the two phase diagrams. In section IV and section V, we study similar properties of this model with OBC in y and x directions separately. Finally, we summarize the main results  and discuss open problems in section VI. 

\section{Model}
First, we consider the following Hermitian model describing nodal line semimetals \cite{yang2019non}:
\begin{equation} 
H_{0}(\bm{k})=(\cos k_{x}+\cos k_{y}+\cos k_{z}-m )\sigma_{x}+\sin k_{z} \sigma_{y},\label{1e}
\end{equation}
where $m=\frac{21}{8}$.  Considering the OBC in z direction,  we can write the Hamiltonian as
\begin{equation}
H(k_{z})=(\cos k_{z}+f_{0}) \sigma_{x}+\sin k_{z} \sigma_{y},\label{2e}
\end{equation}
where $f_{0}=-m+\cos k_{x}+\cos k_{y}$. We diagonalize the Hamiltonian to obtain the PBC energy spectrum and  write down the corresponding Hamiltonian in real space to calculate the OBC energy spectrum. 
\begin{figure}[tbph]
	\centering
	\includegraphics[scale=0.132]{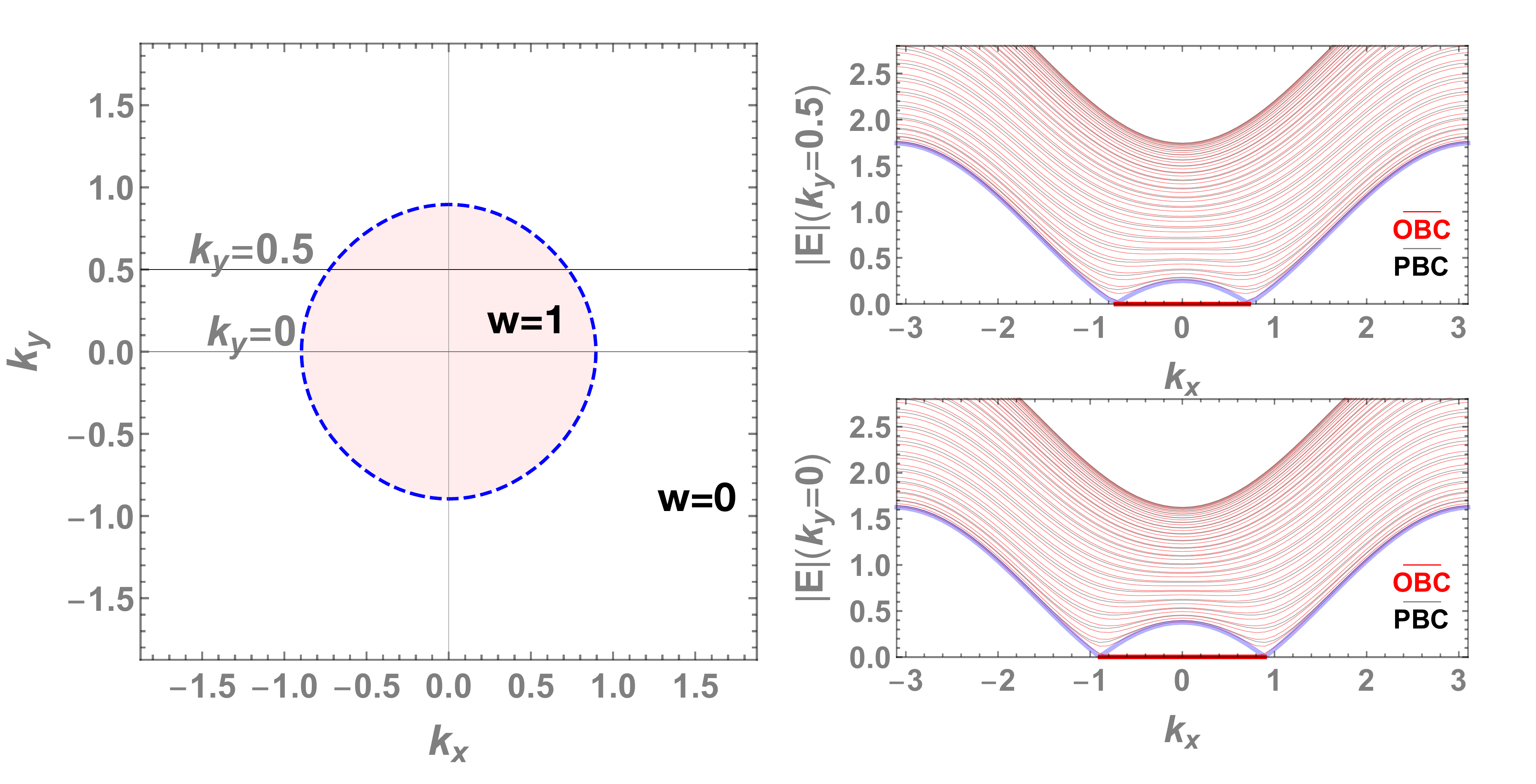}
	\caption{Bulk-boundary correspondence of Hermitian nodal line semimetal. The left subfigure is the phase diagram, in which the pink region represents topological nontrivial phase ($w=1$). The right subfigure presents the PBC and OBC spectrum $|E|-k_{x}$ for $k_{y}=0$ and $k_{y}=0.5$.}
	\label{1}
\end{figure}
\par For convenience, we fix $k_{y}=0$ and $k_{y}=0.5$ to study  $|E|-k_{x}$ relation. Since the Hamiltonian has the chiral symmetry, its eigenvalues have ($E$,$-E$) pairs. So we only need to plot the $|E|-k_{x}$ relation. Shown in Fig.\ref{1}, when $k_{y}=0$ and $k_{y}=0.5$, the PBC spectrum is consistent with OBC spectrum. This shows the bulk-boundary correspondence in this Hermitian model. We  also plot the phase diagram, in which the pink region represents topological nontrivial region with the drumhead surface states, and the rest region represents trivial phase. The boundary of the topological nontrivial phase is determined by the gap-closing condition. It is easy to verify that the boundary is: $\cos k_{x}+\cos k_{y}=\frac{13}{8}$.

\par Now  we add non-Hermitian terms to the Hermitian nodal line semimetal model to generate non-Hermitian semimetals:
\begin{equation}
\begin{aligned}
H&=(\cos k_{z}-m+\cos k_{x}+\cos k_{y}) \sigma_{x}\\
&+\sin k_{z} \sigma_{y}+f_{x}\sigma_{x}+f_{y}\sigma_{y},
\end{aligned}\label{5e}
\end{equation}
where $f_{x}=\frac{i}{2}\sin k_{y}$, $f_{y}=\frac{i}{2}\sin k_{x}$, and $m=\frac{21}{8}$. We don't add terms that contain $\sigma_{z}$ in order to preserve the chiral symmetry. 

\begin{figure}[t]
	\centering
	\includegraphics[scale=0.229]{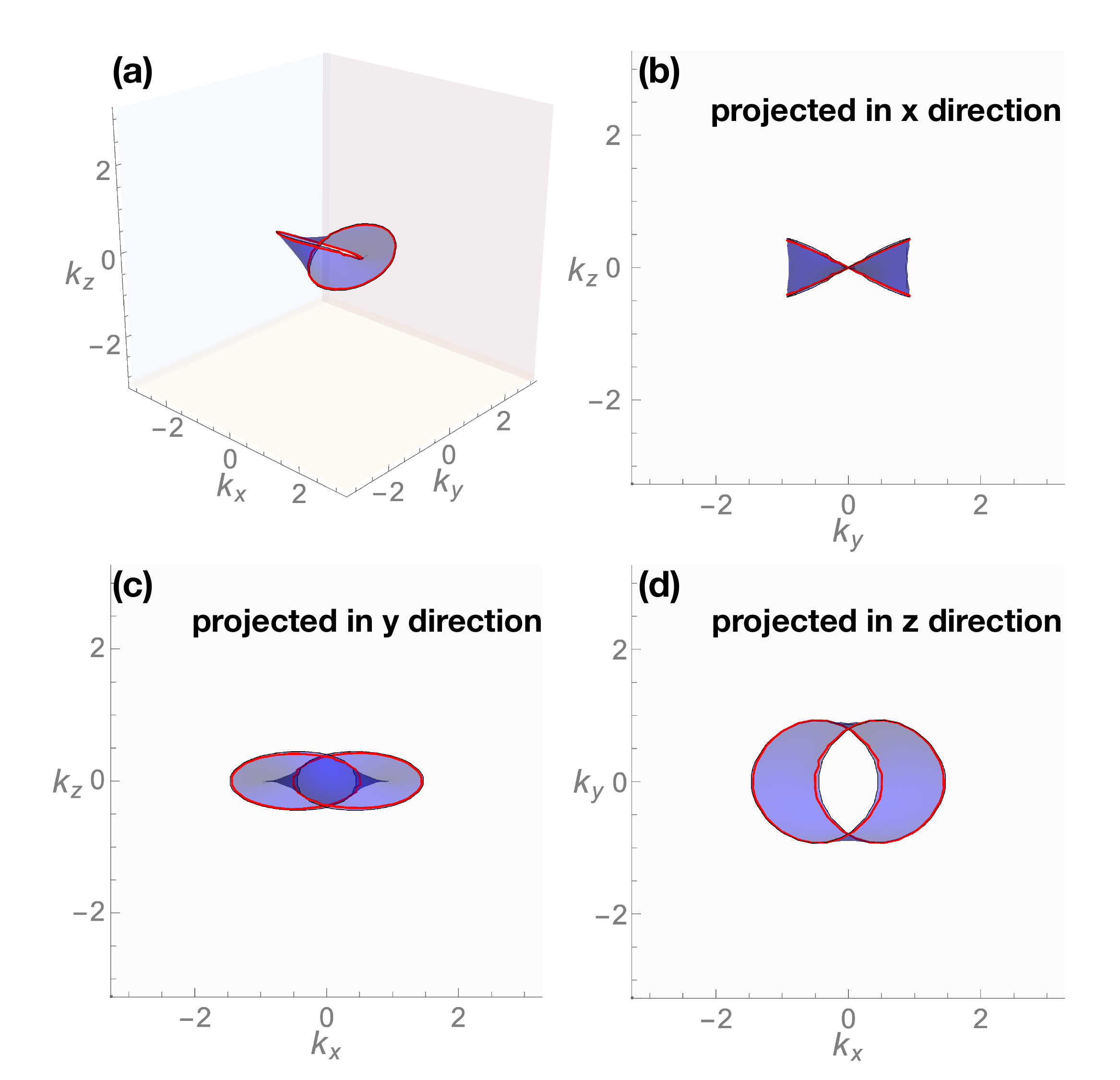}
	\caption{The Fermi surface of the non-Hermitian exceptional line semimetal model: (a) The Fermi surface is colored in blue. The red lines are its boundaries that form a Hopf link; (b) The projection of the Fermi surface in x direction  as two red crossed lines; (c) The projection of the boundary of the Fermi surface in y direction as two closed curves; (d) The projection of the boundary of the Fermi surface in z direction  as two closed curves.}\label{3}
\end{figure}

After diagonalizing the Hamiltonian, we can obtain the eigenvalue equation, i.e. the PBC spectrum 
\begin{equation}
\begin{aligned}
E^2&=(\cos k_{z}-\frac{21}{8}+\cos k_{x}+\cos k_{y}+\\
&\frac{i}{2}\sin k_{y})^2+(\sin k_{z}+\frac{i}{2}\sin k_{x})^2.
\end{aligned}\label{22e}
\end{equation}
By definition,  the Fermi surface of this system requires $Re[E(\bm{k})]=0$, which implies $Re[E^2]<0$ and $Im[E^2]=0$. Thus, the boundary of Fermi surface is $Re[E^2]=0$ and $Im[E^2]=0$. The boundary of the Fermi surface is given by $\sin k_{z}\sin k_{x}+(\cos k_{z}-\frac{21}{8}+\cos k_{x}+\cos k_{y}) \sin k_{y}=0$ and $({\sin k_{z}})^2-\frac{1}{4}(\sin k_{x})^2+(\cos k_{z}-\frac{21}{8}+\cos k_{x}+\cos k_{y})^2-({\sin k_{y}})^2=0$, which is  shown in Fig. \ref{3}(a). 

\par We find that the Fermi surface is a 2D twisting surface. In Fig. \ref{3}(a), the Fermi surface is colored in blue. The boundaries of the Fermi surface are two closed curves colored in red, which form the Hopf-link exceptional lines. We project it in x, y, and z directions to fully present the Fermi surface. As is shown in Fig. \ref{3}(b), the projection of the Fermi surface boundary in x direction consists of two crossed lines. However, both the projections of Fermi surface boundary in y and z directions are two closed curves, as shown in Fig. \ref{3}(c) and Fig. \ref{3}(d) separately. 
\par Next we study the bulk-boundary correspondence of the non-Hermitian semimetal model with OBC in z direction, y direction, and x direction one by one. 

%\par Both numerical method in Ref\cite{yokomizo2019non} and analytic method in  Ref[Yang] to calculate the GBZs apply to 1D models, but the nodal line semimetal model is a 3D model. So in the following text we consider the 3D non-Hermitian nodal line semimetal model with open boundary in x direction, y direction, and z direction separately, which naturally leads to three 1D models. Then we study the bulk-boundary correspondence of these 1D models. 

%{\color{red} It should be emphasized that the boundary state of the 3D non-Hermitian semimetals are determined by the Hamiltonian with all directions with open boundary condition. However, there is no standard method to }

\section{Open Boundary in z Direction}
\par The Hamiltonian for the non-Hermitian semimetal model with open boundary in z direction is:
\begin{equation}
\begin{aligned}
H(k_{z})&=(\cos k_{z}-i\sin k_{z}+p_{z+})\frac{\sigma_{+}}{2}\\
&+(\cos k_{z}+i\sin k_{z}+p_{z-})\frac{\sigma_{-}}{2},\label{55e}
\end{aligned}
\end{equation}
in which $p_{z\pm}=\cos k_{x}+\cos k_{y}-m+\frac{i}{2}\sin k_{y}\pm \frac{1}{2}\sin k_{x}$, $\sigma_{\pm}=\sigma_{x}\pm i\sigma_{y}$, $m=21/8$. The PBC spectra are given by Eq. (\ref{22e}), where $k_{x}$ and $k_{y}$ are variable parameters when we open boundary in z direction.   The OBC spectra can be obtained from the eigenvalues of the corresponding Hamiltonian  as discussed in Appendix (\ref{Appendix C}).
\begin{figure}[htbp]
	\centering
	\includegraphics[scale=0.42]{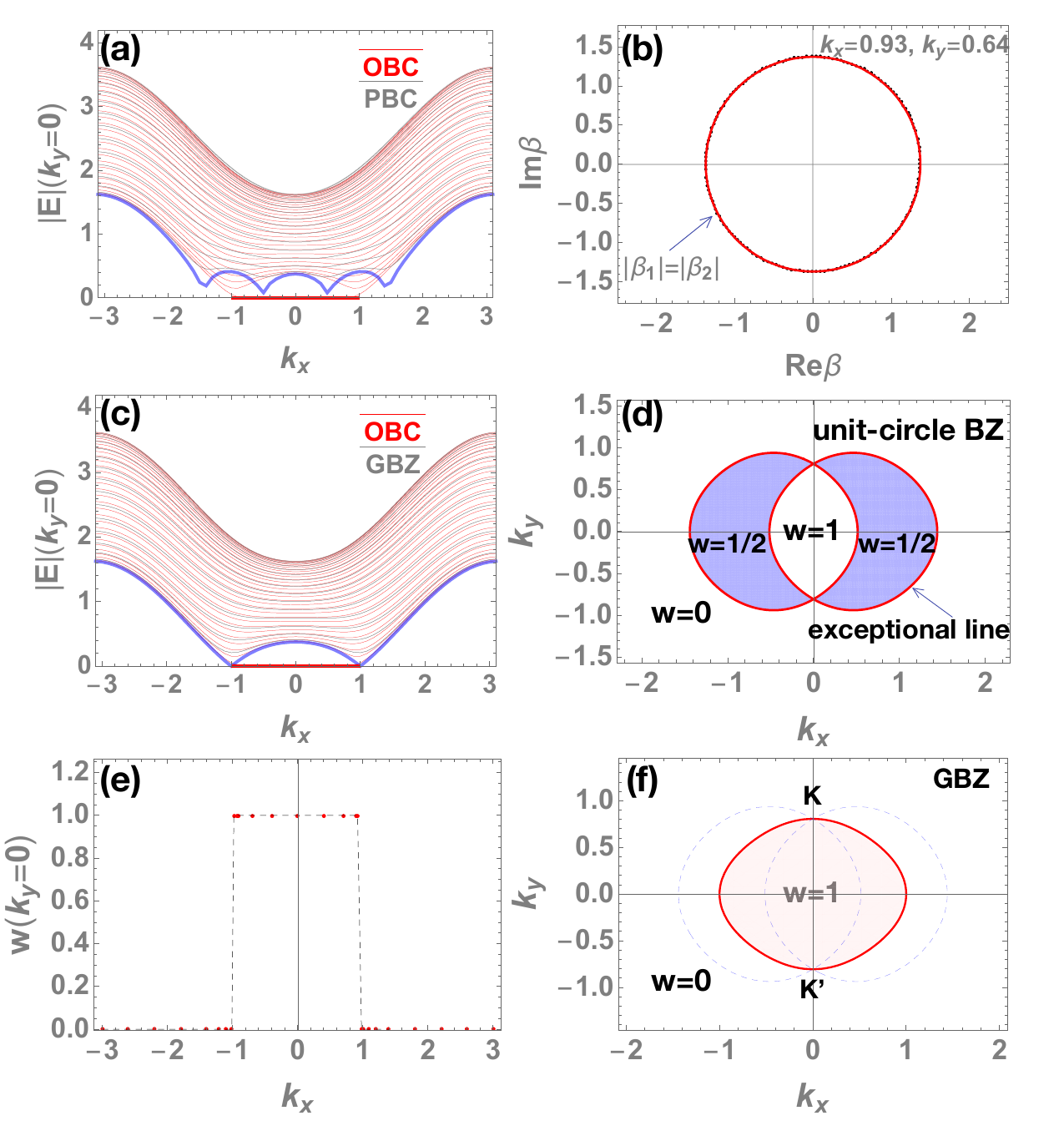}
	\caption{The non-Hermitian exceptional line semimetal model with open boundary in z direction: (a) the significant differences between PBC and OBC spectra; (b) GBZs calculated with numerical method (black dots) and analytic methods (red circle);  (c) the comparison of PBC and OBC spectra ($k_{z}=0$). here the PBC spectra are calculated with GBZs;    (d) the unit-circle BZ phase diagram calculated with PBC and unit-circle BZ; the winding numbers are labelled in each region; the two red closed curves cross the line $k_{y}=0$ at 4 points: $k_{x}=-1.44, -0.51, 0.51, 1.44$. (e) $w-k_{x}$  plotted with $k_{y}=0$; the range of $k_{x}$ corresponding to $w=1$ in Fig. \ref{9}(e) is the same as the  range of $k_{x}$ corresponding to zero modes in Fig. \ref{9}(c); (f) the GBZ  phase diagram with labelled winding number; the pink region corresponds to nontrivial phase with $w=1$, and the rest region is trivial phase with $w=0$; the red exceptional line crosses the line $k_{y}=0$ at $(-1.00,0)$ and $(1.00, 0)$;  the phase boundaries of unit-circle BZ phase diagram in Fig. \ref{9}(d) are drawn as blue dashed lines, which cross the GBZ phase boundary at K and K$^{\prime}$ points.}\label{9}
\end{figure}

\par First, we derive the explicit expression of GBZ and give the parameter region where the bulk-boundary correspondence is broken. Given the matrix form of the real-space Hamiltonian, we can write down the  Schr\"{o}dinger equation in real space: $ \psi_{B,n-1}+(f_{0}+if_{x}+f_{y})\psi_{B,n}=E\psi_{A,n}$ and $(f_{0}+if_{x}-f_{y})\psi_{A,n}+\psi_{A,n+1}=E \psi_{B,n}$.  The solution has the form $(\psi_{A,n},\psi_{B,n})={\beta_{z,n}}(\psi_{A},\psi_{B})$.   The characteristic equation of this model is given by $[(f_{0}+if_{x}+f_{y})\beta_{z}+1][\beta_{z}+(f_{0}+if_{x}-f_{y})]=E^2\beta_{z}$.   The continuum bands condition requires that $|\beta_{z1}|=|\beta_{z2}|$. Applying  Vieta's theorem, we have   
\begin{equation}
|\beta_{z}|=\sqrt{|\frac{f_{0}+if_{x}-f_{y}}{f_{0}+if_{x}+f_{y}}|}.\label{14e}
\end{equation}
Since $f_{y}=\frac{1}{2}\sin k_{x}$,   $|\beta_{z}|=1$ if and only if $k_{x}=0$.  Here we don't consider the case where $k_{x}=\pi$. Thus, the bulk-boundary correspondence is preserved for $k_{x}=0$ but broken for $k_{x}\neq0$. 

\par Next, we present the breakdown of the bulk-boundary correspondence by showing the deviation from unit circles of the GBZs and the discrepancy between PBC spectra and OBC spectra. As an example, we fix $k_{x}= \arccos0.6 \approx0.93, k_{y}= \arccos0.8 \approx0.64$ and  calculate the OBC spectrum. We choose these parameters because it is convenient to apply the analytic method in Ref. \cite{yang2019auxiliary} to calculate GBZ at these parameters. Substituting the OBC spectrum into the characteristic equation and applying the continuum bands condition $|\beta_{z1}|=|\beta_{z2}|$, the trajectory of $\beta_{z1}$ and $\beta_{z2}$ leads to the GBZ shown as discrete black dots in Fig. \ref{9}(b). The red circle in Fig. \ref{9}(b) is the analytic result of GBZ with explicit expression: $(\mathrm{Re}\beta_{z})^2+(\mathrm{Im}\beta_{z})^2=\sqrt{\frac{4369}{1233}}$.  The result is consistent with the numerical one. As we expect, the GBZ is not a unit-circle and the bulk-boundary correspondence is broken when we take $k_{x}=0.93\neq0$. Furthermore,  we fix $k_{y}=0$ and calculate the $|E|-k_{x}$ relations with PBC and OBC.  As shown   in Fig. \ref{9}(a), when $k_{x}\neq0$, there is significant discrepancy between PBC and OBC spectra,  implying the existence of the breakdown of bulk-boundary correspondence.  However, as  shown in Fig. \ref{9}(c), if we substitute $e^{ik_{z}}\rightarrow\beta_{z}$, the PBC spectra are consistent with the OBC spectra. Namely, the introduction of GBZ recovers the bulk-boundary correspondence in this non-Hermitian model. 

\par We then calculate the topological phase diagrams.  When we use a unit-circle Brillouin zone, the gap closing condition $|E|=0$ of Eq. (\ref{22e}) requires both the real part and imaginary part of $E$ to be 0, leading to two equations: $\cos k_{z}+f_{0}\mp f_{y}=0$ and $f_{x}\pm \sin k_{z}=0$. The two equations together give rise to $f_{x}^2+(f_{0}\mp f_{y})^2=1$, which are the red exceptional lines shown in Fig. \ref{9}(d). To specify the topological property of the rest region in $k_{x}-k_{y}$ plane, we calculate winding numbers and label them in each region.   The unit-circle BZ phase diagram is the same as the projection of the Fermi surface in z direction shown in Fig. \ref{3}(d). This is easy to understand, because both the two diagrams are calculated with PBC and unit-circle BZs. To calculate the GBZ phase diagram, we substitute $e^{ik_{z}}$ with $\beta_{z}$ into Eq. (\ref{22e}). The gap-closing condition gives rise to $(f_{0}^2+f_{x}^2-f_{y}^2)^2+4f_{x}^2f_{y}^2=1$, as shown in Fig. \ref{9}(f) where  the topological nontrivial and trivial phases are indicated in the pink and white regions respectively.    To compare the two phase diagrams, the phase boundaries of unit-circle BZ phase diagram in Fig. \ref{9}(d) are drawn as blue dashed curves in Fig. \ref{9}(f), which intersect with the GBZ phase boundary at K and K$^\prime$ points. Fig. \ref{9}(e) shows the $w-k_{x}$ relation with $k_{y}=0$. The range of $k_{x}$ corresponding to $w=1$ in Fig. \ref{9}(e) is consistent with the range of $k_{x}$ for zero modes of OBC spectrum in Fig. \ref{9}(c). Thus,  we can conclude that GBZ phase diagram gives the correct bulk-boundary correspondence. 
\par Next, we explain the differences between the phase boundaries of the two phase diagrams. The phase boundaries in both two phase diagrams are calculated from the gap-closing condition of the characteristic equation:
\begin{equation} 
\begin{aligned}
E^2&=(\frac{1}{2}(\beta_{z}+\frac{1}{\beta_{z}})-m+\cos k_{x}+\cos k_{y}+\\
& \frac{i}{2}\sin k_{y})^2+(\frac{1}{2i}(\beta_{z}-\frac{1}{\beta_{z}})+\frac{i}{2}\sin k_{x})^2. 
\end{aligned}
\end{equation}
 The difference is that the constraint $|\beta_{z}|=1$ is applied to calculate the phase boundary in unit-circle BZ phase diagram while the constraint $|\beta_{z1}|=|\beta_{z2}|$ is applied to obtain the phase boundary in GBZ phase diagram. When we take $|\beta_{z}|=1$, the characteristic equation becomes Eq. (\ref{22e}). This characteristic equation leads to the unit-circle BZ phase diagram. However, when we take $|\beta_{z1}|=|\beta_{z2}|$, the gap-closing condition of the characteristic equation leads to $|\beta_{z}|=\sqrt{|\frac{f_{0}+if_{x}-f_{y}}{f_{0}+if_{x}+f_{y}}|}$. In general, we don't have $|\beta_{z}|=1$. This is the origin of the differences between the phase boundaries of the two phase diagrams.  However, if the constraint $|\beta_{z1}|=|\beta_{z2}|$ leads to $|\beta_{z1}|=|\beta_{z2}|=1$ at some parameters, the characteristic equation will also be Eq. (\ref{22e}). It means that the phase boundaries of unit-circle BZ phase diagram and GBZ phase diagram  will intersect at these parameters.  In our model,  as  $|\beta_{z}|=\sqrt{|\frac{f_{0}+if_{x}-f_{y}}{f_{0}+if_{x}+f_{y}}|}$ and $f_{y}=\frac{1}{2}\sin k_{x}$,   the phase boundaries of the two phase diagrams will intersect at points where $k_{x}=0$. This is clearly shown in Fig. \ref{9}(f), where the two phase boundaries cross at K and K$^{\prime}$ points on the line $k_{x}=0$.
 \par  We may also notice that there exists $\omega=1/2$ region in the unit-circle BZ phase diagram Fig. \ref{9}(d). The $\omega=1/2$ region vanishes in the GBZ phase diagram Fig. \ref{9}(f). According to appendix A, for a two band model $H=R_{+}(\beta)\sigma_{+}+R_{-}(\beta)\sigma_{-}$ with chiral symmetry, the definition of the eigenstate winding number is $\omega=(\omega_{+}-\omega_{-})/2$. $\omega_{+}$ is the winding number of $R_{+}(\beta)$, and $\omega_{-}$ is the winding number of $R_{-}(\beta)$ when $\beta$ goes along the GBZ $C_{\beta}$. We can get $\omega=1/2$ if we choose proper parameters. When we take OBC in z direction of our model Eq. (\ref{1e}) and take the parameters $k_{x}=1.0$ and $k_{y}=0.2$ in Eq. (\ref{22e}), we have $\omega_{+}=1$ and $\omega_{-}=0$, giving rise to $\omega=1/2$. For non-Hermitian systems, we can also define the eigenvalue winding number: $\omega_{E}=(\omega_{+}+\omega_{-})/2$. As is shown in Ref. \cite{zhang2019correspondence}, for OBC spectra and GBZ spectra, we have $\omega_{E}=0$, indicating that $\omega_{+}=-\omega_{-}$. Thus, $\omega_{S}=(\omega_{+}-\omega_{-})/2\in \mathcal{Z}$, and the $\omega_{S}=1/2$ region will vanish in GBZ phase diagram.
\section{Open Boundary in y Direction}
The Hamiltonian for the non-Hermitian  semimetal model with open boundary in y direction is:
\begin{equation}
\begin{aligned}
H(k_{y})&=(\cos k_{y}+\frac{i}{2}\sin k_{y}+p_{y+})\frac{\sigma_{+}}{2}\\
&+(\cos k_{y}+\frac{i}{2}\sin k_{y}+p_{y-})\frac{\sigma_{-}}{2},
\end{aligned}\label{18e}
\end{equation}
in which $p_{y\pm}=\cos k_{x}+\cos k_{z}-m\mp i\sin k_{z}\pm \frac{1}{2}\sin k_{x}$, $m=\frac{21}{8}$. The PBC spectra are given by Eq. (\ref{22e}), where $k_{x}$ and $k_{z}$ are variable parameters when we open boundary in y direction. Besides, we can obtain OBC spectrum by diagonalizing the corresponding real-space Hamiltonian shown in Appendix(\ref{C5}). 
\begin{figure}[t]
	\centering
	\includegraphics[scale=0.41]{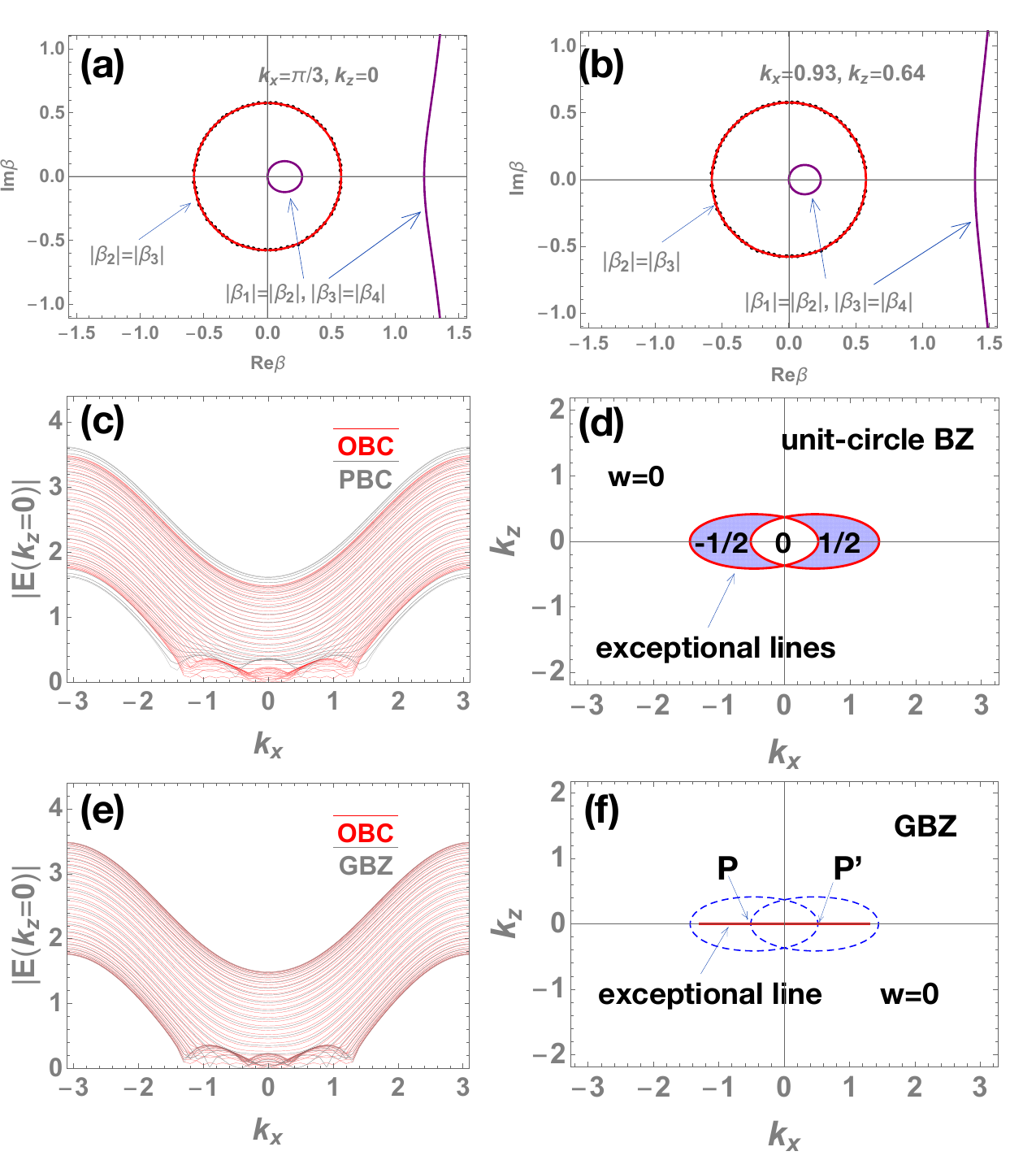}
	\caption{The non-Hermitian nodal line semimetal model with open boundary in y direction:  (a)  the circle GBZ  with $r=\frac{1}{\sqrt{3}}$ when we take $k_{x}=\frac{\pi}{3}, k_{z}=0$; (b) the  circle GBZ with $r=\frac{1}{\sqrt{3}}$ when we take  $k_{x}=0.93, k_{z}=0.64$; (c) the discrepancy between PBC and OBC spectra with $k_{z}=0$ implies that the bulk-boundary correspondence is broken; (d) the boundaries of the PBC phase diagram are two red exceptional lines, intersecting the line $k_{z}=0$ at $k_{x}=-1.44, -0.51, 0.51, 1.44$; (e) the OBC spectrum is consistent with PBC spectrum calculated with GBZ;  (f)  the phase diagram calculated with GBZ; the red line $k_{z}=0, k_{x}\in (-1.29, 1.29)$ is the exceptional line, and the rest region corresponds to the topological trivial phase; the blue dashed lines are the phase boundaries of the unit-circle BZ phase diagram  in Fig. \ref{8}(d), intersecting the red exceptional line at P and P$^{\prime}$.}\label{8}
\end{figure}

\par First, we derive the expression of GBZs and determine the parameter region where the bulk-boundary correspondence is broken.  To calculate the GBZs,  we substitute $e^{ik_{y}}$ in Eq. (\ref{22e}) with $\beta_{y}$ and write down the characteristic equation:
\begin{equation} 
E^2=(\frac{3}{4}\beta_{y}+t_{1}+\frac{1}{4\beta_{y}})^2+{t_{2}}^2\label{25e},
\end{equation} 
where $t_{1}=\cos k_{z}-m+\cos k_{x}$, $t_{2}= \sin k_{z}+\frac{i}{2}\sin k_{x}$, and $m=\frac{21}{8}$. $t_{1}$ and $t_{2}$ are independent of $\beta_{y}$. Then according to Vieta's theorem, the 4 solutions have the relation $|\beta_{a}||\beta_{b}|=\frac{1}{3}$ and $|\beta_{c}||\beta_{d}|=\frac{1}{3}$. Thus, the continuum band condition $|\beta_{y2}|=|\beta_{y3}|$ requires that $|\beta_{y2}|=|\beta_{y3}|=\frac{1}{\sqrt{3}}$. It means that the GBZ is a circle with $r=\frac{1}{\sqrt{3}}\approx0.577$, which is independent of $k_{x}$ and $k_{z}$. We then conclude that the bulk boundary correspondence is broken in this case for any parameters in $k_{x}-k_{z}$ plane. 

\par Next, we manifest the breakdown of the bulk-boundary correspondence by showing  the deviation from unit circles of GBZs and the significant differences between PBC and OBC spectra. For example, we calculate the OBC spectra with parameters $k_{x}=\frac{\pi}{3}, k_{z}=0$, and  substitute the spectra into the characteristic equation Eq. (\ref{25e}). The condition $|\beta_{y2}|=|\beta_{y3}|$ leads to numerical GBZ shown as black dots in Fig. \ref{8}(a). The analytic method in Ref. \cite{yang2019auxiliary} also gives a GBZ. The analytic expression of the GBZ is $x^2+y^2=\frac{1}{3}$, which is plotted as a red circle in Fig. \ref{8}(a).  It is consistent with the numerical result shown in Fig. \ref{8}(a). When we take $k_{x}=0.93, k_{z}=0.64$, we will obtain the circle GBZ with $r=\frac{1}{\sqrt{3}}$ shown in Fig. \ref{8}(b). These results support the fact that the bulk-boundary correspondence is broken for any parameters in the $k_{x}-k_{z}$ plane. To compare the PBC and OBC spectra, we study the $|E|-k_{x}$ relation for $k_z=0$. The result is shown in Fig. \ref{8}(c). The PBC spectra  are quite different from OBC ones, indicating the breakdown of  bulk-boundary correspondence. However, if we apply the transformation $e^{ik_{y}}\rightarrow \beta_{y}=\frac{1}{\sqrt{3}} e^{i\theta}$, the PBC spectra become $
E^2=(\cos k_{z}-21/8+\cos k_{x}+\frac{\sqrt{3}}{2}\cos \theta)^2
+(\sin k_{z}+i\sin k_{x}/2)^2$. The comparison between the OBC spectra and the new PBC spectra calculated with GBZ is shown in Fig. \ref{8}(e). Obviously, the introduction of GBZ in the calculation of PBC spectra recovers the consistency between the PBC and OBC spectra. 

\par The topological phase diagrams can be calculated by applying the gap-closing condition. To obtain the unit-circle BZ phase diagram, we assume that $|\beta_{y}=1|$.  The gap closing condition of PBC spectra Eq. (\ref{22e}) gives red exceptional lines in Fig. \ref{8}(d): $4(\sin k_{z})^2+(\cos k_{x}\pm\frac{1}{2}\sin k_{x}+\cos k_{z}-\frac{21}{8})^2=1$. The winding numbers are labelled in each region in Fig. \ref{8}(d). This phase diagram shows little difference from the projection of Fermi surface in y direction Fig. \ref{3}(c). To get the GBZ topological phase diagram, we substitute $e^{ik_{y}}$ with $ \frac{1}{\sqrt{3}} e^{i\theta}$ in Eq. (\ref{22e}), the gap closing condition leads to the phase diagram shown in Fig. \ref{8}(f). The parameter region for the red exceptional line is $k_{z}=0, k_{x}\in (-1.29, 1.29)$, and the rest region represents topological trivial phase. To show the difference between the two phase diagrams, the phase boundary of the unit-circle BZ phase diagram is shown as blue dashed curves in Fig. \ref{8}(f), which intersect with the red exceptional line of GBZ phase diagram at P and P$^{\prime}$ points.

\par We then analyze the differences between the phase boundaries of the two phase diagrams. The phase boundaries of both the two phase diagrams  are calculated with the gap-closing condition of the characteristic equation Eq. (\ref{25e}). To determine the phase boundary of the GBZ phase diagram, we use the constraint $|\beta_{y2}|=|\beta_{y3}|$ to solve the characteristic equation with $E=0$. The result is $|\beta_{y2}|=|\beta_{y3}|=\frac{1}{\sqrt{3}}\neq1$. Thus, we expect that GBZ phase boundary is different from unit-circle BZ phase diagram. However, we notice that the phase boundaries of the two phase diagrams intersect at P and P$^{\prime}$ points.  In open boundary in z direction case, the characteristic equation is a quadratic equation.  $|\beta_{z1}|=|\beta_{z2}|=1$ leads to the intersecting between the phase boundaries of the two phase diagrams, and vice versa. When there is intersecting between the two phase boundaries at some points, we have $|\beta_{z1}|=|\beta_{z2}|$, and at least one of the solutions satisfy $|\beta_{z}|=1$, leading to $|\beta_{z1}|=|\beta_{z2}|=1$. Nonetheless, it is not the case when we open boundary in y direction. The characteristic equation Eq. (\ref{25e}) is a quartic equation. $|\beta_{y2}|=|\beta_{y3}|=1$ still leads to the intersecting between the phase boundaries of the two phase diagrams. However,  because the characteristic equation has four solutions, at the intersecting points the two conditions $|\beta_{y2}|=|\beta_{y3}|$ and the existence of solution satisfying $|\beta_{y}|=1$  do not necessarily lead to $|\beta_{y2}|=|\beta_{y3}|=1$. In fact, at P$^{\prime}$ point $(k_{x},k_{z})=(0.514,0)$ the characteristic equation Eq. (\ref{25e}) with $E=0$ gives solutions: $\beta_{1}=1,\beta_{2}=0.34 + i0.47,\beta_{3}=0.34- i0.47,\beta_{4}=\frac{1}{3}$ and $|\beta_{1}|=1,|\beta_{2}|=\frac{1}{\sqrt{3}},|\beta_{3}|=\frac{1}{\sqrt{3}},|\beta_{4}|=\frac{1}{3}$. At the intersecting point P$^{\prime}$ the condition $|\beta_{y2}|=|\beta_{y3}|$ is satisfied by the second and third solution, while $|\beta_{y1}|=1$ is satisfied by the first solution. This is the reason why the two phase boundaries intersect at P and P$^{\prime}$ points even if at the two points we have $|\beta_{y2}|=|\beta_{y3}|\neq1$. 
\par  We find that there is a significant difference between y and z open boundary cases. The exceptional line in the GBZ phase diagram in z open boundary case encloses finite area, while the exceptional line in the GBZ phase diagram in y open boundary case is an open arc with zero area. In fact, the behavior of exceptional lines and Fermi surfaces is highly dependent of the concrete property of the non-Hermitian models. When we take open boundary condition in z direction, we get 1D model Eq. (\ref{55e}). We can calculate the OBC spectrum, finding that there are zero modes. The parameter region corresponding to zero modes is consistent with $\omega=1$ region in GBZ phase diagram Fig. \ref{9}(f). Thus, the exceptional line changes from two intersecting closed curves in unit-circle BZ phase diagram into a closed curve enclosing topological nontrivial region in GBZ phase diagram.  We also open boundary in y direction to get 1D model Eq. (\ref{18e}) and calculate the OBC spectrum. However, we cannot find zero modes corresponding to topological nontrivial states. Thus, different from z open boundary case, the exceptional line in y open boundary case collapses into an open arc and doesn't enclose topological nontrivial region with finite area.

\section{Open Boundary in x Direction}
\par The Hamiltonian of the non-Hermitian exceptional line semimetal model with OBC in x direction  is: 
\begin{equation}
\begin{aligned}
H(k_{x})&=(\cos k_{x}+\frac{1}{2}\sin k_{x}+p_{x+})\frac{\sigma_{+}}{2}\\
&+(\cos k_{x}-\frac{1}{2}\sin k_{x}+p_{x-})\frac{\sigma_{-}}{2},
\end{aligned}\label{15e}
\end{equation}
in which $p_{x\pm}=\cos k_{y}+\cos k_{z}-m+\frac{i}{2}\sin k_{y}\mp i\sin k_{z}$, and $m=\frac{21}{8}$. The PBC spectra are given by Eq. (\ref{22e}), where $k_{y}$ and $k_{z}$ are variable parameters when we open boundary in x direction. OBC spectra are calculated from the corresponding real-space OBC Hamiltonians in Appendix (\ref{C4}).

\par First, we  calculate the GBZs and give the parameter region where the bulk-boundary correspondence is broken. To calculate GBZs, we substitute $e^{ik_{x}}$ with $\beta_{x}$ in Eq. (\ref{22e}) to obtain the characteristic equation:
\begin{equation}
\begin{aligned}
E^2&=(\frac{1}{2}(\beta_{x}+\frac{1}{\beta_{x}})+\cos k_{z}-21/8+\cos k_{y}+\\
&\frac{i}{2}\sin k_{y})^2+(\frac{1}{4}(\beta_{x}-\frac{1}{\beta_{x}})+\sin k_{z})^2,
\end{aligned}\label{17e}
\end{equation}
The structure of this characteristic equation is more complex than the  above two cases with open boundary in z and y directions. But we can still get the analytic expression of GBZs when we take $k_{z}=0$. With $k_{z}=0$,  the characteristic equation Eq. (\ref{17e}) is invariant under the transformation $\beta_{x}\rightarrow\frac{1}{\beta_{x}}$. Thus, if $\beta_{x}$ is a solution of the equation, $\frac{1}{\beta_{x}}$ is also  a solution.  Without loss of generality, we suppose that there are four solutions to the equation: $\beta_{1}, \beta_{2},\frac{1}{\beta_{1}}, \frac{1}{\beta_{2}}$, and $|\beta_{1}|\leq|\beta_{2}|\leq1$. The ordering of the moduli of $\beta$s is $|\beta_{1}|\leq|\beta_{2}|\leq|\frac{1}{\beta_{2}}|\leq|\frac{1}{\beta_{1}}|$. Thus, the continuum band condition $|\beta_{2}|=|\beta_{3}|$ requires that $|\beta_{2}|=|\frac{1}{\beta_{2}}|=1$. Thus, the Brillouin zone is a unit circle if $k_{z}=0$. There is no analytic expression of GBZ if $k_{z}\neq0$. However, we can use the theorem in Ref. \cite{zhang2019correspondence} to verify whether the bulk-boundary correspondence is broken. We find that for $k_{z}=0$ the PBC spectra in $Re E-ImE$ plane have winding number $\nu_{E}=0$ with respect to any reference point in the complex energy plane, while for $k_{z}\neq0$ all the PBC spectra have winding number $\nu_{E}\neq0$. According to the theorem in Ref. \cite{zhang2019correspondence}, we can conclude that the bulk-boundary correspondence is preserved for $k_{z}=0$ but broken for $k_{z}\neq0$. 
 \begin{figure}[t]
	\centering
	\includegraphics[scale=0.41]{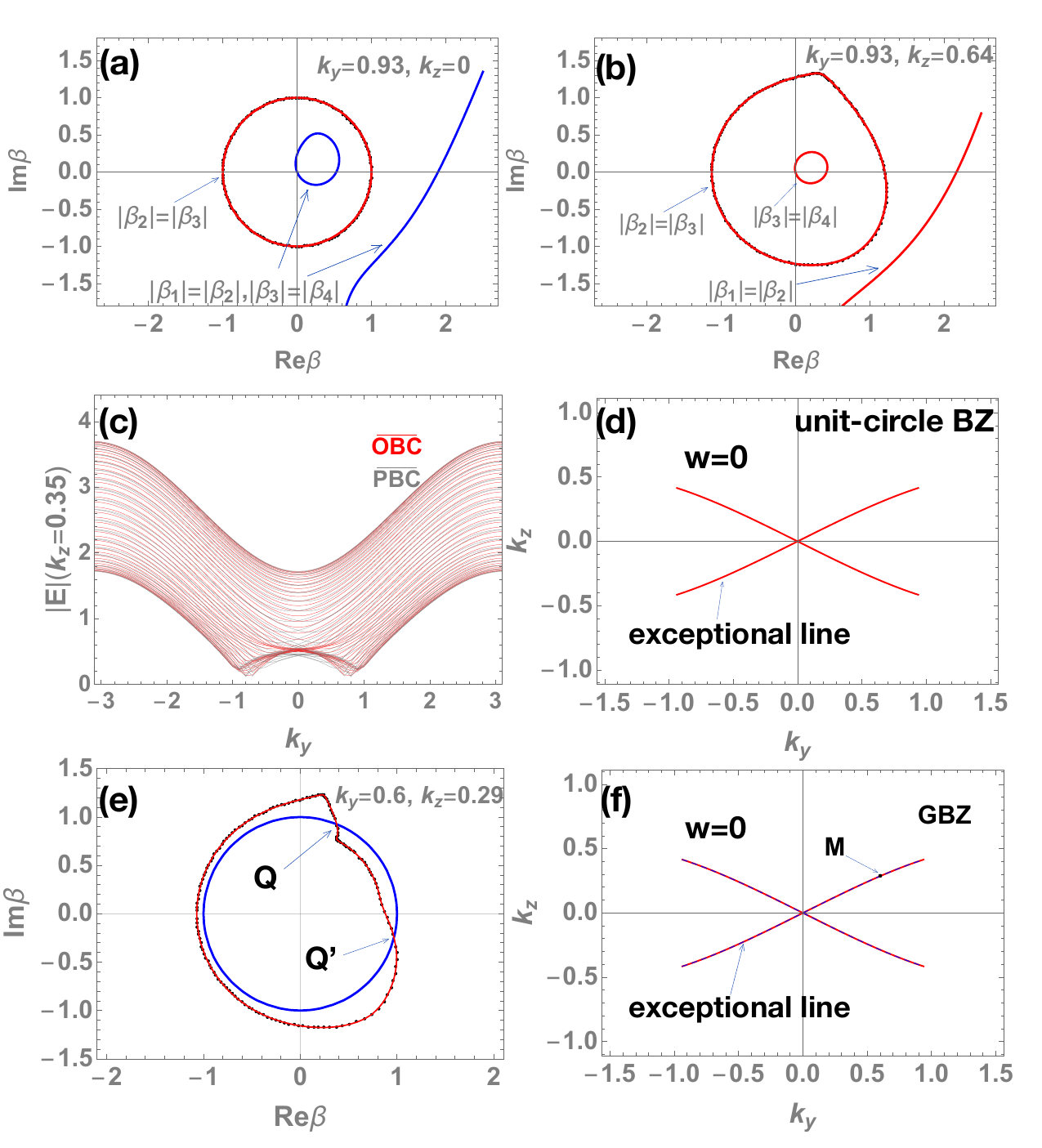}
	\caption{The non-Hermitian nodal line semimetal model with open boundary in x direction:  (a)  the unit circle GBZ with $k_{y}=0.93, k_{z}=0$  from both numerical results (black dots) and analytic results (red closed lines);  (b)  the GBZ with $k_{y}=0.93, k_{z}=0.64$, which is not a unit circle.  (c) the comparison between the PBC and OBC spectra with $k_{z}=0.35$; (d) the unit-circle BZ topological phase diagram with the  end points of the exceptional lines at $(k_{x}, k_{y})=(\pm0.94, \pm0.41)$; (e) GBZ for M point $(k_{y}, k_{z})=(0.6,0.29)$ in which the black dots are numerical results; red lines connect these numerical dots to show the GBZ more clearly; the blue curve is a unit circle that intersect the GBZ at Q and Q$^{\prime}$ points; (f) shows the GBZ phase diagram in which the red lines are the exceptional lines; the dashed blue lines are the exceptional lines of the unit-circle BZ phase diagram.}\label{7}
\end{figure}
\par Next, we show the breakdown of the bulk-boundary correspondence by GBZs and the discrepancy between PBC and OBC spectra. We take $k_{y}=0.93$ and $k_{z}=0$ and calculate the OBC spectrum. Substituting the spectrum into the characteristic equation Eq. (\ref{17e}), the continuum band condition $|\beta_{x2}|=|\beta_{x3}|$ leads to  a unit circle  GBZ  as shown in Fig. \ref{7}(a).  We also take $k_{y}=0.93$, $k_{z}=0.64$. The corresponding GBZ  is not a unit circle as shown in Fig. \ref{7}(b). The two results are  consistent with the previous conclusion that the GBZ is a unit circle if $k_{z}=0$ but is not a unit circle if $k_{z}\neq0$.  Then we study the $|E|-k_{y}$ relation with $k_{z}=0.35$. As is shown in Fig. \ref{7}(c), the differences between PBC and OBC spectra show the breakdown of the bulk-boundary correspondence for $k_{z}\neq0$.
\par  Then we calculate the topological phase diagrams.  To calculate the unit-circle BZ phase diagram, we assume that the Brillouin zone is a unit circle . The gap-closing condition leads to $\cos k_{y}+\cos k_{z}>1.51$ and $\sin k_{z}=\pm\frac{1}{2}\sin k_{y}$, which are two red exceptional lines shown in Fig. \ref{7}(d). The rest region is trivial phase with $w=0$. This phase diagram is the same as the projection of Fermi surface in x direction shown in Fig. \ref{3}(b).  Furthermore, we  calculate the GBZ phase diagram. The gap-closing condition of the characteristic equation Eq. (\ref{17e}) and continuum band condition $|\beta_{x2}|=|\beta_{x3}|$ also lead to $\cos k_{y}+\cos k_{z}>1.51$ and $\sin k_{z}=\pm\frac{1}{2}\sin k_{y}$. The GBZ phase diagram is shown in Fig. \ref{7}(f). To compare the two phase diagrams, we plot the exceptional lines of unit-circle BZ phase diagram as blue dashed lines. 
 We find that the two exceptional lines totally overlap and the two phase diagrams are exactly the same.
\par Finally,  we explain the reasons why the two phase diagrams are identical. Both the two exceptional lines are determined by the gap-closing condition of the characteristic equation Eq. (\ref{17e}).  To obtain the exceptional lines in the GBZ phase diagram, we apply the constraint $|\beta_{x2}|=|\beta_{x3}|$. Then the characteristic equation Eq. (\ref{17e}) with $E=0$ gives rise to $|\beta_{x2}|=|\beta_{x3}|=1$, which is independent of $k_{y}$ and $k_{z}$. This explains why the two phase diagrams are exactly the same.  To understand why we have $|\beta_{x2}|=|\beta_{x3}|=1$ even if $k_{z}\neq0$ and the bulk-boundary correspondence is broken, we choose the $\mathit{M}$ point $(0.6, 0.29)$ in Fig. \ref{7}(f)  on the exceptional line as an example.  To get the GBZ of M point, we fix $k_{y}=0.6, k_{z}=0.29$ and calculate the OBC spectra from the real-space Hamiltonian in Appendix (\ref{C4}).  Substituting the spectra into the characteristic equation Eq. (\ref{17e}), the continuum bands condition $|\beta_{x2}|=|\beta_{x3}|$ leads to the GBZ shown as black dots in Fig. \ref{7}(e). The red lines connect these black dots to present the GBZ more clearly. The GBZ is not a unit circle as we expect.  On the other hand, when we use gap-closing condition $E=0$ and the continuum band condition $|\beta_{x2}|=|\beta_{x3}|$ to solve the characteristic equation, we  have  $\beta_{1}=0.49 + 
i0.22, \beta_{2}=0.38 + i0.93, \beta_{3}=0.97-i0.25, \beta_{4}=0.40-i1.80$ and $|\beta_{1}|=0.54, |\beta_{2}|=|\beta_{3}|=1, |\beta_{4}|=1.84$. $\beta_{2}$ and $\beta_{3}$ are plotted in Fig. \ref{7}(e) as Q and Q$^{\prime}$ points, which are exactly the crossing points of the GBZ and the unit circle. It means that although the GBZ of M point is not a unit circle, it crosses the unit circle at Q and Q$^{\prime}$ points, which correspond to the eigenenergy $E=0$ in the OBC spectrum. This example suggests that it is necessary to introduce GBZ to characterize the bulk-boundary correspondence. Because even if the unit-circle BZ phase diagram is the same as GBZ phase diagram, the bulk-boundary correspondence is broken for $k_{z}\neq0$, which can be clearly shown by the deviation from unit circles of GBZs.

\section{Summary and Discussions}
In summary, we study a non-Hermitian exceptional line semimetal model with open boundary in z, y, and x directions separately. In each case, we calculate the parameter region corresponding to the breakdown of the bulk-boundary correspondence.  The GBZs and the discrepancy between PBC and OBC spectra  present the breakdown of the bulk-boundary correspondence. 
\par We demonstrate that in all considered cases, the numerical method and analytic method  result in the same GBZs. The PBC spectrum calculated with GBZ is consistent with the OBC spectrum. Namely, the introduction of GBZ recovers the bulk-boundary correspondence in these models.  Both unit-circle BZ phase diagrams and GBZ phase diagrams are plotted. The difference between the unit-circle BZ phase diagrams and GBZ phase diagrams highlights the significance of GBZ in characterizing the bulk-boundary correspondence of non-Hermitian models.
\par As is shown in this article, the aforementioned methods to calculate GBZs work well when we open boundary in one direction. However, a systematic method to calculate the GBZs of non-Hermitian models with open boundary in two or more directions hasn't yet been proposed up to now. In future, we hope to find a generalized method to calculate GBZ and define topological invariants for non-Hermitian models with open boundaries in two and more directions.
\section*{ACKNOWLEDGEMENTS}
We thank Kai Zhang sincerely for his helpful suggestions. Jiangping Hu is supported by the Ministry of Science and Technology of China 973 program (Grant No.~2017YFA0303100), National Science Foundation of China (Grant No. NSFC-11888101), and the Strategic Priority Research Program of CAS (Grant No. XDB28000000).

\appendix
\section{Numerical Method to Calculate GBZ}
For a given model with chiral symmetry, we first write it as a Bloch Hamiltonian $H(k)$, then we get $H(\beta)$ through the transformation $e^{ik}\rightarrow\beta$. Next we solve the characteristic equation $\det[H(\beta)-E]=0$. Supposing that there are $2M$ degrees of freedom in this equation for given $E$, we can get 2M solutions:
 \begin{equation}
|\beta_{1}|\leq |\beta_{2}|\leq \cdots \leq|\beta_{2M}|.\label{A1}
\end{equation}
In principle, $E$ and GBZ can be got through the condition $|\beta_{M}|=|\beta_{M+1}|$   \cite{yokomizo2019non}. We calculate OBC spectrum and get eigenvalues $E$s for given parameters. Then we can substitute $E$s into the characteristic equation $\det[H(\beta)-E]=0$ and get $2M$ solutions $\beta$s. GBZ $C_{\beta}$ is given by the condition:$|\beta_{M}|=|\beta_{M+1}|$. 
\par The definition of winding number is also put forward for non-Hermitian systems: $w=\frac{i}{2\pi}\int_{C_{\beta}}Tr[q^{-1}(\beta)dq]$, where $C_{\beta}$ is the GBZ, and $q$ is a submatrix of $Q$ matrix \cite{yokomizo2019non}. In particular, for a two band model:
\begin{equation}
\mathcal{H}_{\beta}=R_{+}(\beta)\sigma_{+}+R_{-}(\beta)\sigma_{-}
=\left(
\begin{matrix}
0&R_{+}(\beta)\\
R_{-}(\beta)&0 
\end{matrix}
\right),\label{A2}
\end{equation}
the winding number is defined \cite{yokomizo2019non} as 
\begin{equation}
w=-\frac{1}{2\pi}\frac{[\arg R_{+}(\beta)-\arg R_{-}(\beta)]_{C_{\beta}}}{2}.\label{A3}
\end{equation}
In this definition, the winding number is $\omega=(\omega_{+}-\omega_{-})/2$ with $\omega_{+}=-[\arg R_{+}(\beta)]_{C_{\beta}}/(2\pi)$, $\omega_{-}=-[\arg R_{-}(\beta)]_{C_{\beta}}/(2\pi)$. The solutions of the characteristic equation $\beta$s form the GBZ, which is a closed curve $C_{\beta}$. $R_{+}(\beta)$ and $R_{-}(\beta)$ map $\beta$ to two closed curves, having winding numbers $\omega_{+}$ and $\omega_{-}$ separately. 
\section{Analytic Method to Calculate GBZ}
Ref. \cite{yang2019auxiliary} makes use of the mathematical tool resultant to get so-called auxiliary generalized Brillouin zone (aGBZ).  GBZ can be extracted from aGBZs. Let's consider a model with characteristic equation $f(\beta, E)=0$. Supposing that the highest order of the poles of the characteristic equation is $p$, the condition for GBZ is $|\beta_{p}|=|\beta_{p+1}|$. But the direct application of this condition is intractabe. Thus, we relax the condition to be $|\beta_{j}|=|\beta_{j+1}|$, so the requirement for GBZ becomes $f(\beta, E)=f(\beta e^{i\theta},E)=0$. 
\par Obviously, there are 5 variables $\beta_{x},\beta_{y},E_{x},E_{y},\theta$ ($\beta_{x}$ represents $Re \beta$ while $\beta_{y}$ represents $Im \beta$; $E_{x}$ represents $Re E$ while $\beta_{y}$ represents $Im \beta$). We know that the Brillouin zone is an equation that contains only $\beta_{x}$ and $\beta_{y}$, thus we need eliminate $E_{x}, E_{y}, \theta$. First we calculate the resultant $R^{f,f^{\theta}}(\beta)$ to eliminate $E$, which is the resultant between $f(\beta, E)$ and $f^{\theta}(\beta, E)$. To eliminate $\theta$, we need to calculate the resultant (we denote it as $F_{\alpha}(\beta_{x}, \beta_{y})$) between the real part and imaginary part of $R^{f,f^{\theta}}(\beta)$. Finally, the algebraic equation 
\begin{equation}
F_{\alpha}(\beta_{x}, \beta_{y})=0.\label{B1}
\end{equation} gives aGBZs, and we can then apply the condition $|\beta_{p}|=|\beta_{p+1}|$ to get real GBZs. It is noteworthy that this method can give the analytic expression for GBZs, so we needn't consider the lattice size or suffer from numerical errors.
\section{The OBC Real-space Hamiltonians} \label{Appendix C}
\par The model with OBC in z direction: under the basis  $\Psi=(C_{A1},C_{B1},C_{A2},C_{B2},C_{A3},C_{B3},\cdots)^{T}$ where $C^{\dagger}_{Ai}$ creates an electron on the A site of $i$th unit cell in z direction, the real-space Hamiltonian in matrix form is:
\begin{equation}
\left(
\begin{matrix}
0&f_{0}+if_{x}+f_{y}&0&\cdots\\
f_{0}+if_{x}-f_{y}&0&1&\cdots\\
0&1&0&\cdots\\
\cdots&\cdots&\cdots&\ddots
\end{matrix}
\right),\label{C3}
\end{equation} where $f_{0}=\cos k_{x}+\cos k_{y}-m$, $f_{x}=\frac{1}{2}\sin k_{y}$, $f_{y}=\frac{1}{2}\sin k_{x}$, and $m=\frac{21}{8}$.  

\par The model with OBC in y direction: under the basis $\Psi=(C_{A1},C_{B1},C_{A2},C_{B2},C_{A3},\cdots)^{T}$ where $C^{\dagger}_{Ai}$ creates an electron on the A site of $i$th unit cell in y direction,  the corresponding Hamiltonian in real space is:
\begin{equation}
\left(
\begin{matrix}
0&q_{1y}+q_{2y}&0&\cdots\\
q_{1y}-q_{2y}&0&\frac{1}{2}+\frac{1}{4}&\cdots\\
0&\frac{1}{2}-\frac{1}{4}&0&\cdots\\
\cdots&\cdots&\cdots&\ddots
\end{matrix}
\right),\label{C5}
\end{equation} where $q_{1y}=\cos k_{x}+\cos k_{z}-m$, $q_{2y}=-i\sin k_{z}+\frac{1}{2}\sin k_{x}$, and $m=\frac{21}{8}$.

\par The model with OBC in x direction: with the basis $\Psi=(C_{A1},C_{B1},C_{A2},C_{B2},C_{A3},\cdots)^{T}$  where $C^{\dagger}_{Ai}$ creates an electron on the A site of $i$th unit cell in x direction, the corresponding real-space Hamiltonian in matrix form is:
\begin{equation}
\left(
\begin{matrix}
0&q_{1x}+q_{2x}&0&\frac{1}{2}-\frac{i}{4}&\cdots\\
q_{1x}-q_{2x}&0&\frac{1}{2}+\frac{i}{4}&0&\cdots\\
0&\frac{1}{2}+\frac{i}{4}&0&q_{1x}+q_{2x}&\cdots\\
\frac{1}{2}-\frac{i}{4}&0&q_{1x}-q_{2x}&0&\cdots\\
\cdots&\cdots&\cdots&\cdots&\ddots
\end{matrix}
\right),\label{C4}
\end{equation} where $q_{1x}=\cos k_{z}-m+\cos k_{y}+\frac{i}{2}\sin k_{y}$, $q_{2x}=-i \sin k_{z}$, and $m=\frac{21}{8}$.

\bibliography{NHrefLINK.bib}
\bibliographystyle{apsrev4-1}
\end{document}